\documentclass[journal=jacsat,manuscript=article]{achemso}
\usepackage[version=3]{mhchem} 
\usepackage{comment}
\usepackage{multirow} 
\usepackage{array} 
\usepackage{makecell} 
\usepackage{amsmath}
\usepackage{xr-hyper}
\usepackage{hyperref}
\usepackage{ragged2e}
\usepackage[none]{hyphenat}
\usepackage{gensymb}
\providecommand{\tb}[1]{\textbf{(#1)}}

\newcommand{\im}{\ensuremath{\mathrm{Cu_6Sn_5}}}
\newcommand{\stannate}{\ensuremath{\mathrm{Sn(OH)_6^{2-}}}}
\newcommand{\stannite}{\ensuremath{\mathrm{Sn(OH)_3^-}}}

\newcommand{\ksn}{\ensuremath{\mathrm{K_2Sn(OH)_6}}}

\SectionNumbersOn
\author{Sofia K. Catalina}
\affiliation[]{Department of Materials Science \& Engineering, Stanford University, United States}
\altaffiliation{These authors contributed equally to this work.}

\author{Kyle Frohna}
\affiliation{Department of Materials Science \& Engineering, Stanford University, United States}
\alsoaffiliation{Department of Electrical Engineering, Stanford University, United States}
\altaffiliation{These authors contributed equally to this work.}

\author{Willow Thompson}
\affiliation[]{Department of Materials Science \& Engineering, Stanford University, United States}

\author{Katherine Harmon}
\affiliation[]{Department of Materials Science \& Engineering, Stanford University, United States}

\author{Dasol Yoon}
\affiliation[]{Department of Materials Science \& Engineering, Stanford University, United States}

\author{Jianbo Wang}
\affiliation[]{Department of Materials Science \& Engineering, Stanford University, United States}

\author{Colin Ophus}
\affiliation[]{Department of Materials Science \& Engineering, Stanford University, United States}

\author{Daniel N. Congreve}
\affiliation[]{Department of Electrical Engineering, Stanford University, United States}
\author{William C. Chueh}
\affiliation[]{Department of Materials Science \& Engineering, Stanford University, United States}
\alsoaffiliation[Stanford University]{Department of Energy Science \& Engineering, Stanford University, United States.}
\alsoaffiliation[SLAC National Laboratory]{Applied Energy Division, SLAC National Laboratory, United States.}
\title[]{Visualizing Degradation in Anode-Free High-Utilization Aqueous Batteries Across Cell Lifetime}

\abbreviations{}

\begin{document}
\vspace{-20pt}
\hspace{100pt} {\href{mailto:congreve@stanford.edu}{congreve@stanford.edu} \hspace{10pt}  {\href{mailto:wchueh@stanford.edu}{wchueh@stanford.edu}} 


\begin{abstract}
\noindent\textit{Operando} microscopy has unveiled key mechanistic insights in battery materials during early cycling, but long-term characterization to unveil material evolution, degradation, and failure remain limited. 
To address this gap, we develop a custom \textit{operando} optical microscope that captures images across hundreds of cycles and hours using optically accessible, anode-free pouch cells. 
We image through-plane, bulk-representative electrodeposition behavior of aqueous tin metal anodes, which are promising due to their high energy density but whose reactivity limits practical cycle life. 
We show that substrate governs the morphology and stability of plated tin, particularly at high plated capacities. 
Specifically, copper substrates exhibit a multi-stage tin growth mode, which results in high overpotentials and irreversible active material loss at high plated capacities. In contrast, graphite substrates display a single-stage growth mode with slower kinetics. 
Using this insight, we balance performance and stability to demonstrate a high-utilization (70\%, 630 mAh g$^{-1}_{Sn}$) porous graphite substrate Sn anode with high efficiency and long lifetime.
Our results underscore the importance of material and device optimization guided by \textit{operando} characterization across device lifetime with broad applicability to electrochemical systems.

\end{abstract}
\section{Introduction} \label{sec:intro}

The parasitic reactions and degradation processes that limit battery performance occur heterogeneously, over long timescales, and across many cycles. Gaining a full understanding of the performance and degradation of a battery chemistry requires the ability to measure the cell with high spatial resolution and statistical power across the entire device lifetime. 
While powerful, \textit{ex-situ} and \textit{in-situ} microscopy only record isolated snapshots of a battery during its lifetime\cite{louliDiagnosingCorrectingAnodefree2020,bondSituImagingElectrode2022}, while \textit{operando} battery microscopy studies are generally limited to initial cycles\cite{limOriginHysteresisLithium2016,wangStructuralDistortionInduced2020}, meaning the pathways leading to cell degradation often remain unknown. 
Optical microscopy is low-cost, parallelizable, and has been successfully applied to intercalation electrodes in Li/Na ion cells\cite{merryweatherOperandoOpticalTracking2021,merryweatherOperandoMonitoringSingleparticle2022,luMultiscaleDynamicsCharging2023,luUnravellingElectrochemomechanicalProcesses2025,gopalMappingOutFast2026} and plating/stripping of metals in organic\cite{sanchezPlanViewOperandoVideo2020,sanchezLithiumStrippingAnisotropic2021} and aqueous\cite{liTailoringMetalElectrode2022,xuHighlyReversibleTin2024,zhangConstructingDenseMorphology2025} electrolytes. 
However, capturing extended duration operation across many cycles remains a challenge due to the difficulty of engineering \textit{operando} optical cells with electrochemical behavior equivalent to conventional cells\cite{pujariIdentifyingCurrentCollectors2024,pujariOptimalCoinCell2026}, with most experiments being performed in modified geometries to allow surface-down or side-view imaging\cite{maDynamicInterfacialStability2021,duObservationZnDendrite2023,baiTransitionLithiumGrowth2016,woodDendritesPitsUntangling2016}.
Further, additional engineering efforts are necessary to enable an optical system with sufficient resolution and stability to produce high-fidelity images across cell lifetime.

To capture relevant \textit{in-situ} behavior, we design and implement an \textit{operando} optical microscope to image battery operation and degradation across hundreds of cycles and hours and apply this to optically accessible pouch cells with identical performance to opaque equivalents. We apply this methodology to anode-free, metal electrodeposition batteries as they represent a promising pathway to high energy density. Across chemistries, their primary limitation lies in the intrinsically dynamic and heterogeneous nature of metal deposition, which drives morphological instability and poor reversibility. In aqueous systems, these challenges are compounded by low metal-ion solubility, parasitic reactions, and interfacial instability which limit achievable energy density and cycle life\cite{wangStabilizationPerspectiveMetal2021,juEnergeticAqueousBatteries2022,wangHighlyReversibleZinc2018,hongHighlyTexturedMetal2025}. \textit{Operando} characterization across cell lifetime is critical to building a mechanistic understanding of these complex dynamics.

Tin (Sn) is a compelling chemistry to overcome these constraints. 
In alkaline electrolytes, Sn exhibits high hydrogen evolution reaction (HER) overpotential, non-dendritic morphologies, and high theoretical capacity (903.1 mAh g$^{-1}_{Sn}$, 257.3 Ah L$^{-1}$) via the 4e$^-$ reaction\cite{zhangChallengesOpportunitiesRechargeable2025}. 
During cycling, \stannate{} reversibly deposits onto the substrate as metallic Sn via a \stannite{} intermediate (Fig. \ref{fig:fig1}c).

\begin{equation}
    \mathrm{\stannate + 2e^- \rightarrow \stannite + 3 OH^-}
\end{equation}
\begin{equation}
    \mathrm{\stannite + 2e^- \rightarrow Sn + 3 OH^-}
\end{equation}

Electrodeposition reactions have high theoretical capacities, but morphology, utilization of active material, and reversibility are governed by the local interfacial environment, which dictates nucleation kinetics, growth modes, and parasitic reactivity.
For example, recent work has largely focused on the kinetically favorable Sn/Sn(II) chemistry despite the fact that the reaction is fundamentally limited in capacity (451.5 mAh g$^{-1}_{Sn}$) in alkaline media\cite{luSynergisticLigandEngineering2026,tanSnIIPyrophosphateComplexLow2025,wangFlexibleSnAirBatteries2025,xuTendencyRegulationCompeting2023,zhouHighEnergySnNi2023} and is severely constrained in volumetric energy density (10.7 Ah L$^{-1}$) due to low Sn(II) solubility in acidic media\cite{chenStudyHighCurrent2015,weiInvestigationAqueousRechargeable2019,zengNovelTinbromineRedox2019,ouyangDendriteFreeSnAnode2020,zhouElucidatingEffectsComponent2020,liStaticTinManganese2023,zhangHighcapacitySnMetal2023,changLowAcidityChlorideElectrolyte2024,yuHighlyReversibleTin2024,leeDynamicsMetalAnode2025,xiaoHighlyRechargeableAqueous2025,yuHighArealCapacity2025,zhangElectrodepositingTexturedSn2025,lanActivatingProgressiveSn22026, xuHighlyReversibleTin2024,zhangConstructingDenseMorphology2025}.
Furthermore, the substrate is not a passive current collector but an active determinant of electrodeposition behavior \cite{manfredjordanElectrodepositionTinIts1995,gamburgTheoryPracticeMetal2011}.
Substrate effects are evident across the literature. 
Carbonaceous electrodes, particularly graphite, are widely used in both flow\cite{zhouElucidatingEffectsComponent2020,chenStudyHighCurrent2015,zengNovelTinbromineRedox2019,zhouSnFeFlowBattery2018,luSynergisticLigandEngineering2026,tanSnIIPyrophosphateComplexLow2025,yangUreaInducesUniform2025,yaoDendriteFreeTinAnode2021} and static\cite{ouyangDendriteFreeSnAnode2020,xuHighlyReversibleTin2024,weiInvestigationAqueousRechargeable2019,wangReversibleFourelectronSn2024} configurations and consistently exhibit island-like Sn nucleation, high initial overpotential, and weak adhesion due to unfavorable interfacial interactions.
In contrast, copper (Cu) substrates appear to demonstrate more favorable deposition behavior. Strong Sn–Cu interactions and intermetallic formation lower nucleation barriers, promote uniform growth, and improve coulombic efficiency (CE), establishing Cu as the prevailing benchmark for Sn anodes \cite{lanActivatingProgressiveSn22026,yuHighlyReversibleTin2024,zhangHighcapacitySnMetal2023,xuTendencyRegulationCompeting2023,wangFlexibleSnAirBatteries2025}. We likewise have observed enhanced bidirectional kinetics and interfacial alloying on Cu, enabling efficient 4e$^-$ cycling\cite{wangCuSubstrateBidirectional2025}.
Despite this consensus, nearly all static demonstrations are restricted to low utilization of available Sn inventory, low areal capacity, and limited calendar life (see Figs. \ref{fig:litcellformat}–\ref{fig:litsubformat}), raising the question of whether the apparent superiority of Cu persists under the high-utilization conditions and calendar lifetimes relevant to practical Sn batteries.

In this work, we deploy \textit{operando} microscopy to systematically compare both porous and planar graphite and Cu substrates for alkaline Sn anodes across a wide range of plating capacities and cycles, linking interfacial evolution to morphology and performance in optically transparent pouch cells. 
We find that although Cu exhibits favorable kinetics at low areal capacities, its reactivity with Sn drives progressive interfacial degradation at higher areal plated capacities, leading to losses in both coulombic and voltaic efficiency.
By contrast, graphite substrates, while kinetically slower, promote uniform and stable Sn deposition at high areal plated capacities.
These findings reveal a fundamental tradeoff between substrate kinetics and interfacial stability: while enhanced kinetics enable favorable early-cycle behavior of Cu substrates, carbon-based substrates' inertness under high-utilization conditions is essential for long lifetime. This framework enables stable aqueous Sn anodes with high specific and plated areal capacities over hundreds of cycles and hours. More broadly, our results highlight the essential role of \textit{operando} characterization in capturing interfacial evolution under practically relevant conditions over extended timescales.




\section{Main} \label{sec:results}
\subsection{\textit{Operando} Microscopic Visualization of Multi-Stage Tin Anode Plating}
\label{Cumicroscopysection}
To characterize the alkaline Sn metal anode, we fabricated single-layer, transparent pouch cells and demonstrated highly-reproducible, long-duration cycling experiments with equivalent performance to aluminum pouch cells (Figs. \ref{fig:fig1}a-b). Cells were assembled in the discharged state, where the anode active material was completely dissolved in the 2.4 M \ksn{} + 0.4 M KOH anode electrolyte. To probe the Sn anode's morphological evolution during cycling, we built a custom, wide-field optical microscope schematically shown in Fig. \ref{fig:fig1}c, which provides bright-field illumination to the sample with a white light-emitting diode (LED) and images the light scattered onto a complementary metal-oxide semiconductor (CMOS) camera. We used a 20X magnification, 0.75 NA oil-immersion objective lens and a 300 mm focal length tube lens for a total system magnification of 30X with diffraction-limited lateral resolution of $\sim$430 nm. 
We employed active focus stabilization to keep the plane of interest in sharp focus using a feedback loop and demonstrated stability across hundreds of hours and cycles.

With high-fidelity cell and microscope engineering, we directly captured surface evolution across cell lifetime. 
The combination of transparent pouch material and porous substrates enables direct visualization of Sn plating and stripping from behind the current collector in a cross-plane geometry, identical to their layout in opaque pouch cells without modifying the cell architecture.
Visualizing behavior in a high performance, realistic cell architecture is essential, as electrodeposition morphology is strongly governed by cell design.

Cycling porous Cu substrate pouch cells under \textit{operando} optical microscopy, we tracked the coupled evolution of plated Sn and the underlying substrate during cycling. Fig. \ref{fig:fig1}d shows the electrochemical response, and Figs. \ref{fig:fig1}e-n present images of the discharged and charged states across five cycles, with a high charge capacity of 632 mAh g$^{-1}_{Sn}$ and 5.8 mAh cm$^{-2}_{geo}$ corresponding to 70\% utilization of Sn each cycle. 
We also observed the effect of extended cycling on the Cu substrate at 15\% Sn utilization over 150 cycles in an \textit{operando} experiment in the same cell configuration (Fig. \ref{fig:PC2250_manycycles}).
At high utilization, pronounced cycle-to-cycle evolution of Sn morphology was evident, with both mossy growth and large faceted particles emerging. Given the high electrolyte mobility\cite{chenMathematicalModelingExperimental2026b}, we attribute these variations in morphology to local changes in lattice strain at the Sn-Cu interface. Notably, we observed continuous changes to the coloration and morphology of the Cu substrate.  
We hypothesize that this color change is caused by a combination of semi-irreversible Cu-Sn alloy formation (Fig. \ref{fig:fig3}) and surface roughening which decreases specular reflections (Fig. \ref{fig:fig2}). This high degree of reactivity is inconsistent with an ideal inert substrate and has important implications for cell stability. 

\begin{figure}[h!]
  \centering
  \includegraphics[width=0.8\textwidth]{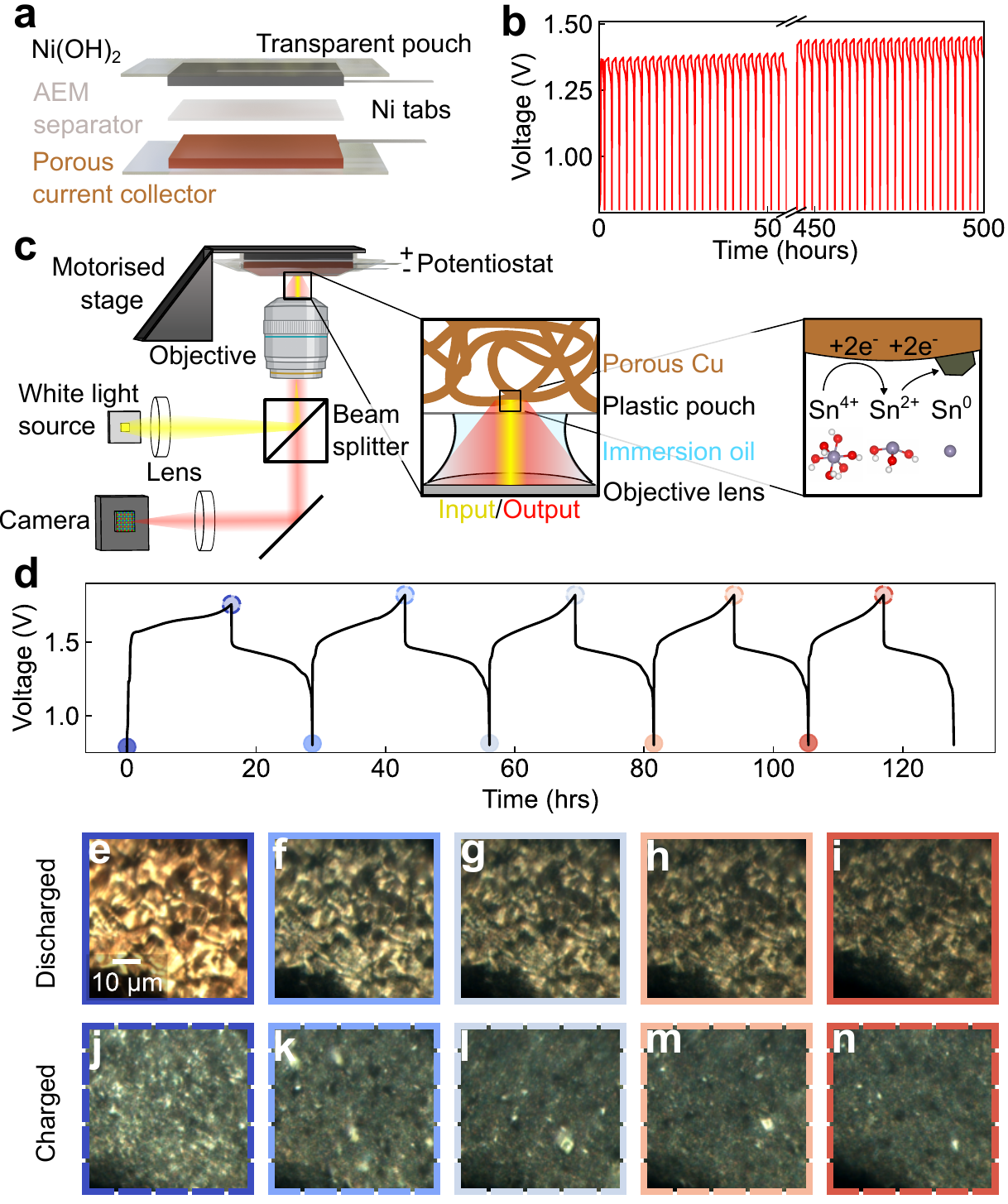}
  \caption{\tb{a} Schematic of the \textit{operando} pouch cell. \tb{b} Galvanostatic cycling curves of a Sn-anode, Cu substrate pouch cell. \tb{c} Schematic of the \textit{operando} optical microscopy setup and the Sn anode reactions taking place at the Cu substrate. \tb{d} Galvanostatic cycling of Cu foam pouch cell at high (70\%) utilization taken during an \textit{operando} microscopy experiment. \tb{e-i} Optical images of the Cu substrate in discharged state at the start of each cycle marked in d. \tb{j-n} Optical images of the plated Sn in the charged state at the middle of each cycle marked in d.}
  \label{fig:fig1}
\end{figure}

\subsection{\textit{Operando} Characterization of Cu Substrate Degradation}
\label{Cudegradationsection}

To further understand the mechanisms captured by \textit{operando} optical microscopy, we probed the Cu substrate and Sn morphology evolution during charge. The voltage curve (black) in Fig. \ref{fig:fig2}a shows a galvanostatic charge of a pouch cell to 70\% utilization. We observed an initial small plateau at $\sim$1.2 V, followed by charge plateaus at $\sim$1.57 V and $\sim$1.67 V. 
The two-step character of the main charge plateau is clearly resolved in the differential capacity (dQ/dV) profile shown in Fig. \ref{fig:fig2}b. 
The emergence of two distinct features in the derivative plot indicates a transition in kinetics during plating.
We captured images of the Cu substrate throughout Sn plating and extracted grayscale intensity values plotted in red in Fig. \ref{fig:fig2}a. 
Images extracted at the points marked on the charge curve in Fig. \ref{fig:fig2}a are plotted in Fig. \ref{fig:fig2}c-g. By combining the electrochemical data, the images and their grayscale intensities, and the X-ray diffraction results in the following section, we directly visualize a three-stage Sn plating mechanism. 
\\
In Stage 0 (Fig. \ref{fig:fig2}a and c), the optical intensity and voltage both increase before plateauing in tandem, corresponding to the reduction of Cu$_2$O. 
This behavior is a first cycle activation process involving the native oxide species present on the Cu surface (Fig. \ref{fig:pristineSEMXRD}). These species are not regenerated during standard cycling unless Cu oxidation is induced during over-discharge (Fig. \ref{fig:overdischargeSEMXRD}).
In Stage 1 (Fig. \ref{fig:fig2}d), the optical intensity increases as the electrochemistry reaches the main plating plateau. This increase in brightness is attributed to Sn nucleation on the Cu surface and rapid formation of a thin \im{} intermetallic layer. 
The formation of this interface was characterized in the lead-free soldering literature\cite{sunPropertiesMicrostructuresSnAgCuX2015,zhangNewChallengesMiniaturization2020} and reported in our previous work \cite{wangCuSubstrateBidirectional2025}. 
The mobile Cu ions diffuse quickly into the deposited Sn to form a brighter-colored \im{} layer, further characterized using scanning transmission electron microscopy energy dispersive X-ray spectroscopy (STEM-EDX) of a Sn deposit on a Cu substrate in Figs. \ref{fig:fig3}a-c and e. We identify a $\sim$0.5 $\mu$m region of compositional intermixing and Cu diffusion along Sn grain boundaries. 
Our \textit{operando} measurements highlight that this reaction occurs rapidly during early stage electrodeposition (Fig. \ref{fig:low_utXRD}) and further characterization using scanning electron microscopy (SEM) suggests that intermetallic growth proceeds throughout plating (Figs. \ref{fig:fig3}d, \ref{fig:10pct_crossecSEM}-\ref{fig:70pct_crosssecSEM}). 
Stage 2 growth (Fig. \ref{fig:fig2}e) corresponds with the first sharp peak at $\sim$1.57 V in Fig. \ref{fig:fig2}b. A thin layer of mossy Sn grows across the Cu surface, resulting in a sharp decrease in brightness.
Finally, once all preferred sites have been covered, likely governed by the interfacial energies of the polycrystalline Cu surface (Fig. \ref{fig:pristineSEMXRD}), Stage 3 plating proceeds at a higher overpotential in conjunction with the second peak at 
$\sim$1.67 V (Fig. \ref{fig:fig2}f-g).
The increase in charge overpotential coincides with an increase in optical intensity and marks the transition from initial mossy growth to three-dimensional island formation, consistent with a Stranski–Krastanov growth mechanism \cite{stranskiZurTheorieOrientierten1937,teichertSelforganizationNanostructuresSemiconductor2002}. 
Additional \textit{operando} studies of the Cu substrate behavior are found in Figs. \ref{fig:PC1960_singlecycle}-\ref{fig:PC1960_multicycle}.

To probe how these changing growth modes affect cell stability, we compared identical pouch cells cycled for 500 hours at 5\% (Stages 0-2) and 70\% Sn utilization (Stages 0-3). Choosing a time point rather than cycle number allowed us to compare varied utilization while retaining equivalent total amounts of Sn plated and stripped (approximately 125 mAh capacity throughput), with the only difference being the plated Sn capacity per cycle.
Fig. \ref{fig:fig2}h shows galvanostatic cycling curves of a 5\% utilization pouch cell during the first cycle (solid line) and after 500 hours of cycling (dashed line). The first cycle CE is low due to a fraction of the charge going to the formation of the intermetallic and Cu oxide reduction. 
After 500 hours of cycling, the CE has increased substantially, with concomitant increases in average charge/discharge potentials causing substantial improvement in battery performance (Fig. \ref{fig:Cu-porous-5pct}), consistent with previous reports\cite{wangCuSubstrateBidirectional2025}. However, at 70\% utilization, we see dramatically different electrochemical behavior at 500 hours (Figs. \ref{fig:fig2}i,\ref{fig:Cu-porous-70pct}). The charge overpotential rapidly increases until the cell reaches a noisy plateau at $\sim$2 V which we attribute to HER, and a plateau emerges at $\sim$0.8 V on discharge. Taken together, these measurements clearly demonstrate that a combination of intermetallic formation and Stage 3 growth causes irreversible battery degradation.


\begin{figure}[h!]
  \centering
  \includegraphics[width=0.8\textwidth]{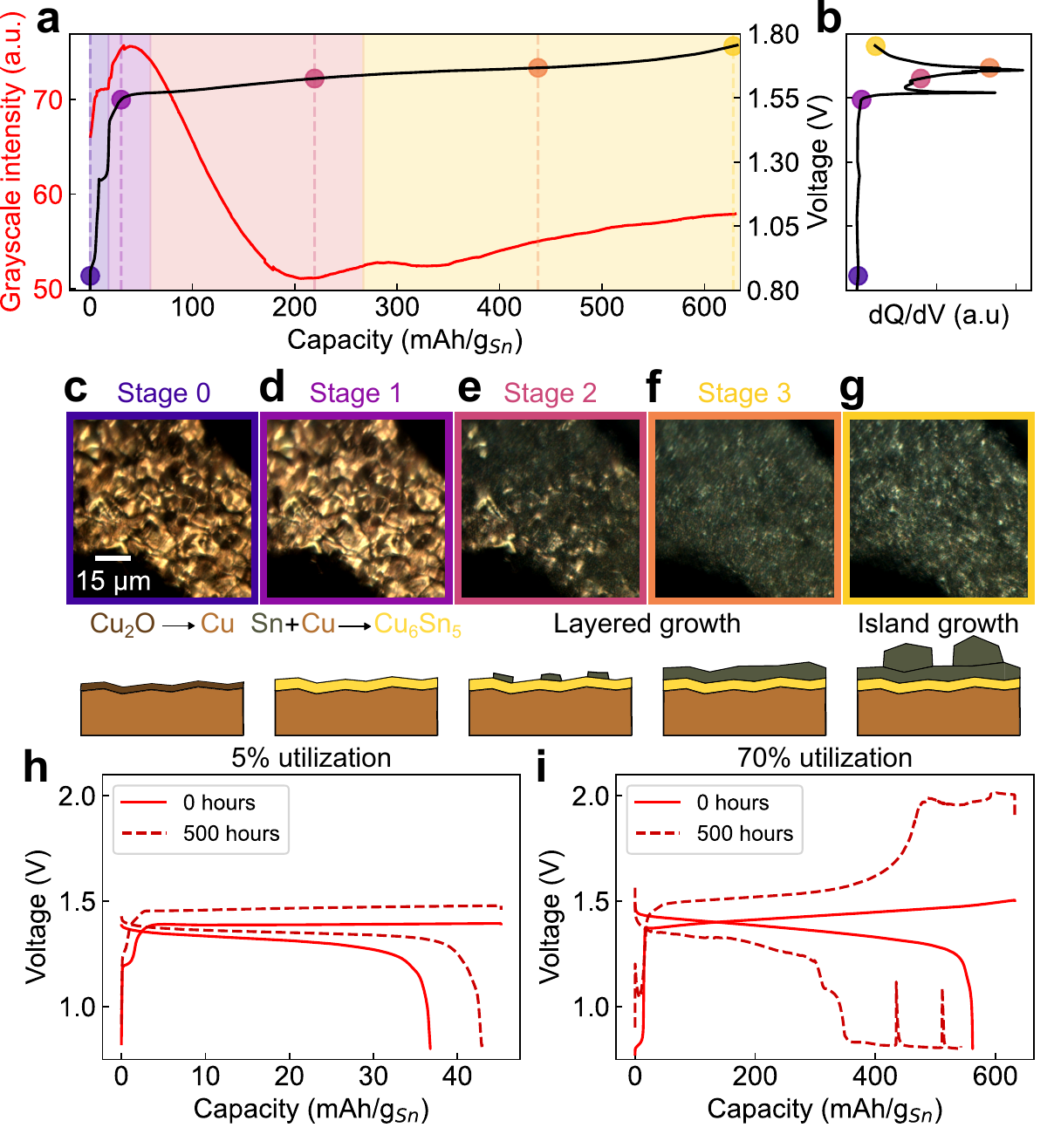}
  \caption{\tb{a} Voltage (black) and grayscale optical intensity (red) versus capacity for a porous Cu substrate, Sn anode battery. The grayscale intensity value is extracted from the regions imaged in panels (c)-(g). Shaded regions correspond to the oxidation stage and three growth stages shown in panels (c)-(g). \tb{b} dQ/dV plot extracted from the data in panel (a). Panels \tb{c}-\tb{g} show optical microscopy images of a region of the porous Cu substrate during Sn plating along with schematics of the reactions occurring at each stage. Panel \tb{c} shows the pristine Cu surface with native Cu oxides. Panel \tb{d} shows the now reduced Cu surface that has brightened in color due to the reduction of the Cu oxide and the formation of a thin \im{} intermetallic layer. \tb{e} and \tb{f} show the progressive growth of a thin layer of mossy Sn across the surface of Cu. \tb{g} shows the growth mode of the Sn change from layered growth to an island morphology. Galvanostatic charge-discharge curves of pristine (solid lines) and cycled (dashed lines) porous Cu substrate pouch cells cycling at \tb{h} low (5\%) and \tb{i} high (70\%) utilization.}
  \label{fig:fig2}
\end{figure}

Next, we probed the \im{} reversibility by performing \textit{ex-situ} grazing incidence X-ray diffraction (GIXRD) on planar Cu substrates charged in pouch cells to low (5\% and 10\%), intermediate (30\% and 50\%) and high (70\%) utilizations and discharged to 0.8 V (Fig. \ref{fig:fig3}f). 
After one low utilization cycle, only Cu and a small amount of Cu$_2$O are observed, suggesting reversible intermetallic formation and potentially small amounts of Cu oxidation at low plated capacity ($\le$ 0.64 mAh cm$^{-2}$). 
This is further supported by SEM images of these substrates showing a pristine surface with faceted Cu$_2$O particles (Fig. \ref{fig:2DCu_Cyc1SEM}a-c). 
At intermediate utilizations, metallic Sn and \im{} phases emerge, accompanied by a dramatic roughening of the surface (Fig. \ref{fig:2DCu_Cyc1SEM} d-e). This roughening rationalizes the continuous darkening of the Cu substrate surface as a function of cycling, as seen in Fig. \ref{fig:fig1} as the roughening reduces specular reflections and is consistent with observations of lead-free solders, where rapid Cu diffusion forms a rough, highly-featured phase boundary\cite{sunPropertiesMicrostructuresSnAgCuX2015,zhangNewChallengesMiniaturization2020}.
Finally, at high utilization, a new set of Bragg peaks indexed to SnO emerges in tandem with further surface roughening (Fig. \ref{fig:fig3}f, Fig. \ref{fig:2DCu_Cyc1SEM}). This suggests that at significant plated capacity, it becomes kinetically competitive to oxidize the intermetallic to SnO during discharge rather than de-alloy to metallic Sn and Cu.
These results validate our \textit{operando} optical finding of a capacity-dependent transition in intermetallic behavior: as plated Sn capacity increases, Cu–Sn alloy formation becomes progressively less reversible, driving active material loss, surface roughening, and higher charge overpotentials that could accelerate parasitic HER.

Motivated by the appearance of metal oxide peaks (Fig. \ref{fig:fig3}f) and the low voltage discharge plateau (Fig. \ref{fig:fig2}i) in the 70\% utilization samples, we performed intentional over-discharge experiments to 0.2 V on both pristine and cycled Cu substrates (Fig. \ref{fig:overdischargeSEMXRD}). Over-discharging a pristine cell yields a small capacity plateau near $\sim$0.4 V, which arises from direct oxidation of the Cu surface to Cu$_2$O as confirmed by ex-situ SEM and GIXRD (Fig. \ref{fig:overdischargeSEMXRD}b--d).
After ten cycles at 10\% utilization, the over-discharge behavior changes markedly: the plateau shifts to higher voltage ($\sim$0.6 V) and carries significantly more capacity (Fig. \ref{fig:overdischargeSEMXRD}a). 
SEM reveals complete surface coverage by micron-scale particles, and GIXRD identifies them as a mixture of Cu$_2$O, SnO$_2$, and residual unoxidized \im{}, indicating that oxidation of the intermetallic results in substantial roughening of the substrate surface.
Returning to the long-duration experiments in Figs. \ref{fig:fig1} and \ref{fig:PC2250_manycycles}, the gradual changes in voltaic efficiency, optical signal, and plating morphology over hundreds of hours are consistent with the irreversible growth of this intermetallic layer and its cumulative impact on cell performance.
Taken together, these results show that \im{} formation promotes competing side reactions that degrade both efficiency and cycle life under high-utilization conditions, and that Cu substrate stability is a critical determinant of long-term Sn anode durability.
                                            
\begin{figure}[h!]
  \centering
  \includegraphics[width=0.75\textwidth]{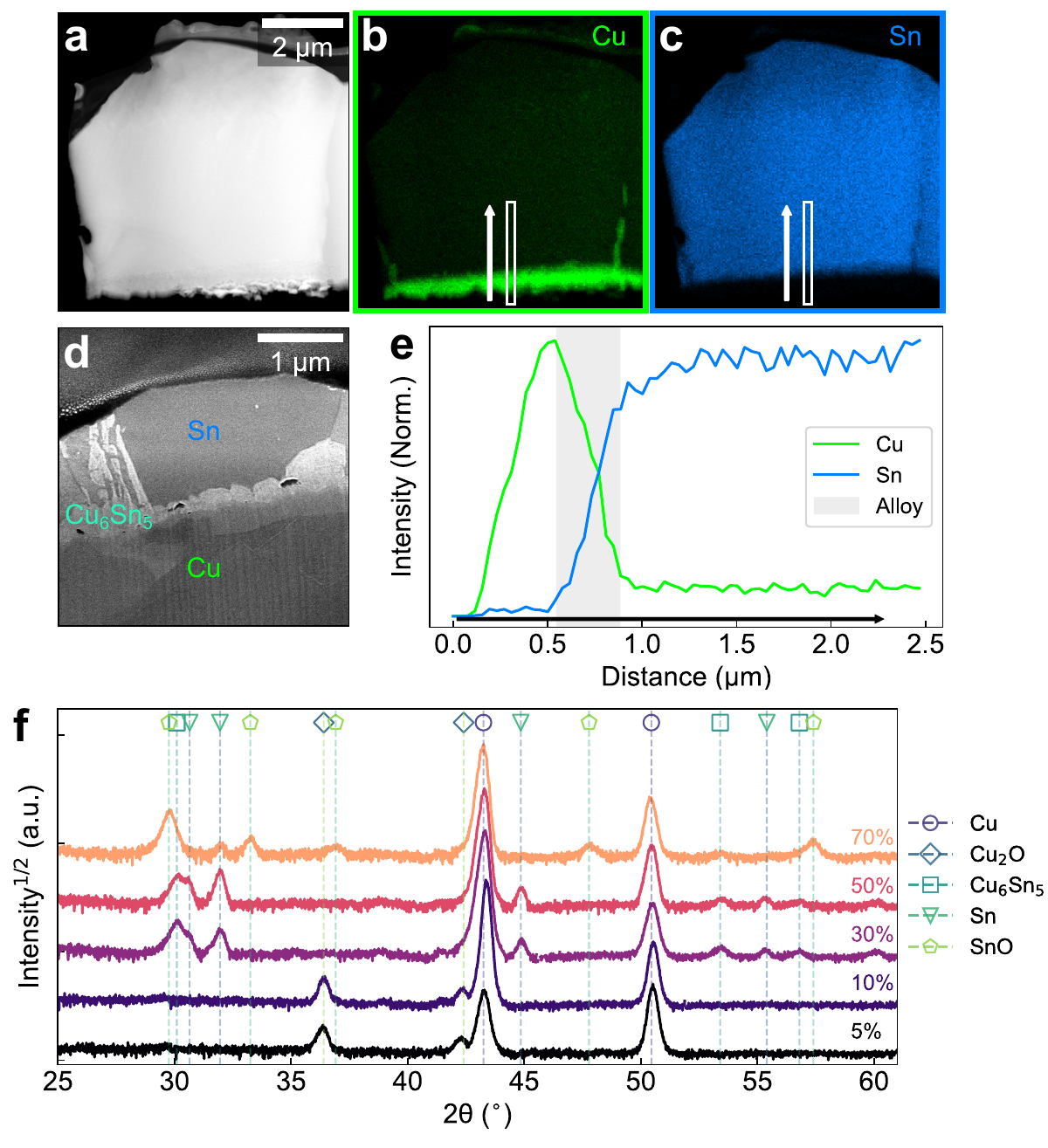}                                            
  \caption{\tb{a} HAADF STEM image, \tb{b} Cu K$_{\alpha}$ EDX image and  \tb{c} Sn L$_{\alpha}$ EDX image of a Sn grain grown on a Cu substrate with pixel-size limited resolution of 3.85 nm. \tb{d} Cross sectional SEM image of a Cu substrate after a 70\% utilization charge. \tb{e} Line cut of Cu K$_{\alpha}$ (green) and Sn L$_{\alpha}$ X-ray intensity extracted from the boxes in panels b and c in the direction of the white arrows. \tb{f} Grazing incidence X-ray diffraction patterns (Cu K$_{\alpha}$ source) of discharged Cu substrates as a function of their charge utilization. Vertical dashed lines indicate the highest intensity peaks from the present crystalline phases.} 
  \label{fig:fig3}
\end{figure}

\subsection{\textit{Operando} Microscopic Visualization of Sn Plating on Graphite Substrates}

In Fig. \ref{fig:fig4}a, we show the charge curve and grayscale optical intensity from a graphite felt pouch cell during an \textit{operando} experiment and a dQ/dV curve exhibiting a single plateau in Fig. \ref{fig:fig4}b. Figs. \ref{fig:fig4}c-g show \textit{operando} microscopy images of the growth of Sn particles on graphite fibers during charge at the points marked in Fig. \ref{fig:fig4}a. The plating capacity per real area for the graphite cells was set to be identical to the porous Cu cells by adjusting electrolyte volume to account for differences in the effective substrate area. 

There are two key differences in the behavior of the graphite substrate as compared to Cu.
First, Sn grows in a single-stage mechanism where Sn nuclei form and grow into isolated faceted tetragonal or quasi-spherical crystallites that increase in size with continued deposition (Fig. \ref{fig:fig4}c-g).
Second, there is no observable change to graphite morphology following discharge due to the inert nature of the surface (Fig. \ref{fig:PC1839_multicycle}).
Figs. \ref{fig:fig4}h-i demonstrate the performance of graphite pouch cells for their first cycle and after 500 hours of cycling at 5\% and 70\% utilization, respectively.
At 5\% utilization, charge is largely unaffected between 0 and 500 hours (Fig. \ref{fig:gr-porous-5pct}), while the distinct two-plateau discharge behavior observed on graphite fibers due to sluggish stannite (\stannite{}) oxidation kinetics improved over cycling. 
This is likely due to enhanced wetting of the porous graphite over time, allowing more facile diffusion of \stannite{} back to the surface for subsequent oxidation.
After 500 hours at 70\% utilization cycling, the end of charge exhibits a sharp increase in voltage corresponding to the onset of HER, while the discharge voltage profile does not change significantly after extended cycling (Figs. \ref{fig:fig4}i, \ref{fig:gr-porous-70pct}). 
Notably, the disproportionation of \stannite{} occurring during discharge at increased utilizations elongates the upper discharge plateau (SI note \ref{suppnote-disprop}), thus increasing voltaic efficiency (VE).

\begin{figure}[h!]
  \centering
  \includegraphics[width=0.75\textwidth]{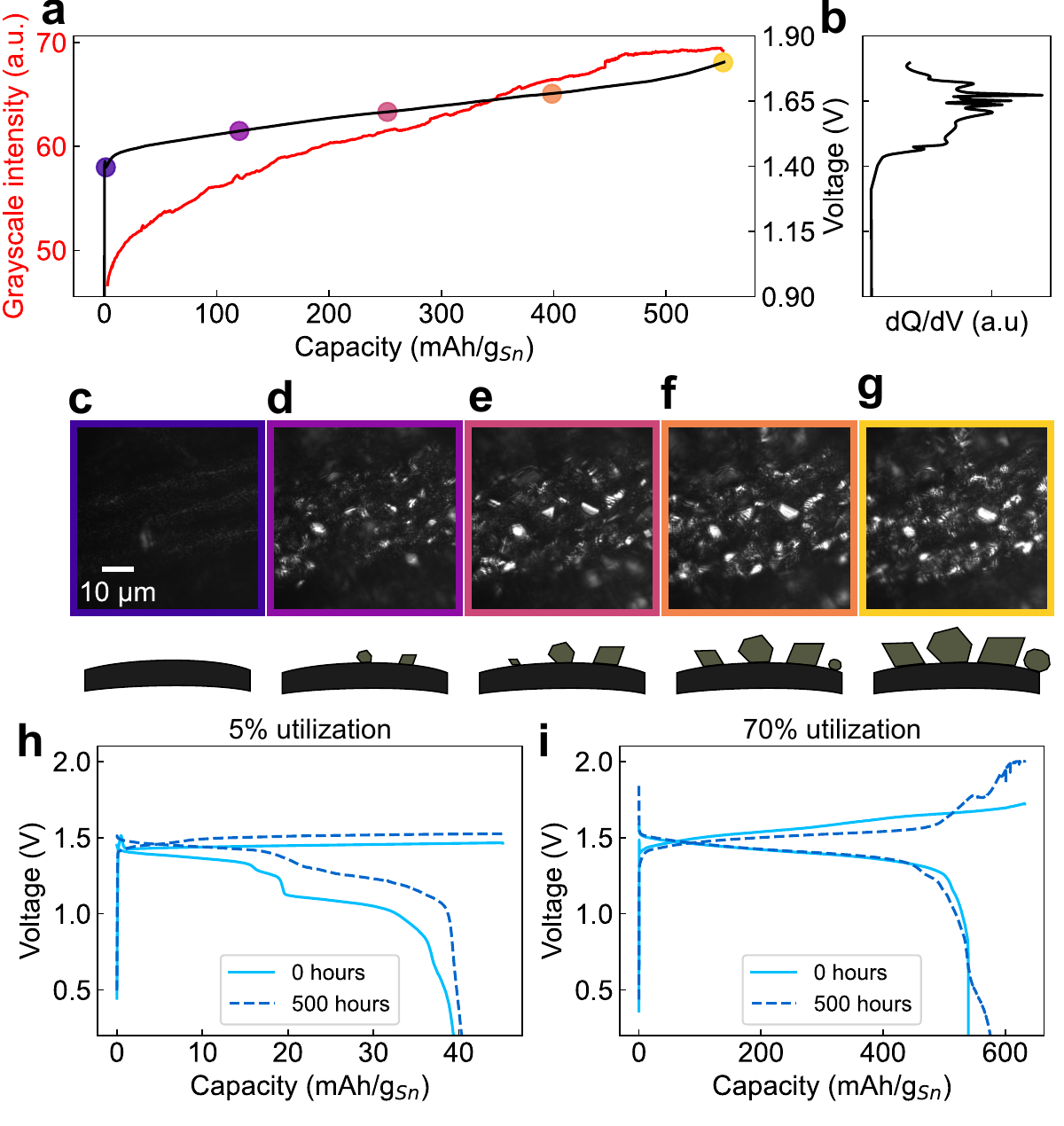}
  \caption{\tb{a} Voltage (black) and grayscale optical intensity (red) versus capacity for a porous graphite substrate, Sn anode battery. \tb{b} dQ/dV plot extracted from the data in panel (a). Panels \tb{c}-\tb{g} show optical microscopy images of a region of the porous graphite during Sn plating along with schematics of the reactions occurring at each stage. Galvanostatic charge-discharge curves of pristine (solid lines) and cycled (dashed lines) porous graphite substrate pouch cells cycling at \tb{h} low (5\%) and \tb{i} high (70\%) utilization.}
  \label{fig:fig4}
\end{figure}

\subsection{High Performance Sn-Anode Batteries}
\label{planarsection}

To summarize our findings, we compare the performance of porous Cu and graphite substrate batteries as a function of utilization. Figs. \ref{fig:Cu-porous-5pct}-\ref{fig:gr-porous-70pct} show representative voltage profiles and performance over 500 hours. 
We plot the CE and VE of these cells as a function of utilization (Fig \ref{fig:fig5}a-b).
At utilizations below 50\%, porous Cu substrates clearly out-compete graphite substrates in both CE and VE, demonstrating $\sim$95\% CE for Cu cells, while graphite cell CE remains limited to $\sim$90\%.
However, due to progressive degradation of the Cu-Sn interface discussed in Section \ref{Cudegradationsection}, at 70\% utilization, Cu substrate CE drops to 83\% and is outperformed by its graphite counterpart.
The same trends are reflected in VE, where at low utilizations Cu outperforms graphite, but at high utilizations, high plating overpotential on Cu and the activity of the disproportionation reaction bring the two substrates into parity. 
The improved VE and sustained CE of the graphitic substrate at high utilization suggest that although Cu substrates offer superior reaction kinetics at beginning of life, the enhanced reversibility of Sn deposition on graphite ultimately makes it the viable platform for efficient, high-capacity, long cycle-life Sn anodes.

Our \textit{operando} optical microscopy has revealed that the multi-stage growth and degradation behavior of Sn on Cu is strongly surface coverage dependent (Fig. \ref{fig:fig1}). 
To further validate these mechanistic findings, we limit surface area by cycling pouch cells with planar Cu and graphite substrates, thereby increasing surface coverage and effective Sn thickness at a given utilization. 
It follows that the transition to island growth will occur at lower utilizations for a planar Cu substrate, exacerbating degradation due to the increased surface coverage.
Figs. \ref{fig:cu-planar-5pct}-\ref{fig:gr-planar-70pct} confirm that while graphite performance remains stable at high areal capacities of Sn, planar Cu demonstrates a far earlier onset of island growth corresponding with reduced CE and VE even at low utilizations (Fig. \ref{fig:SIplanarsub}).
In Figs. \ref{fig:fig5}c-d, CE and VE are plotted against plated Sn capacity per real unit area, accounting for the high surface area of the porous substrates. 
Graphite CE remains consistent at all plated capacities, though at higher utilizations VE is improved as increased activity of the disproportionation reaction circumvents graphite's sluggish stannite oxidation discharge reaction.
However, Cu degrades progressively with increasing areal capacity, consistent with our mechanism. 

\begin{figure}[h!]
  \centering
  \includegraphics[width=0.75\textwidth]{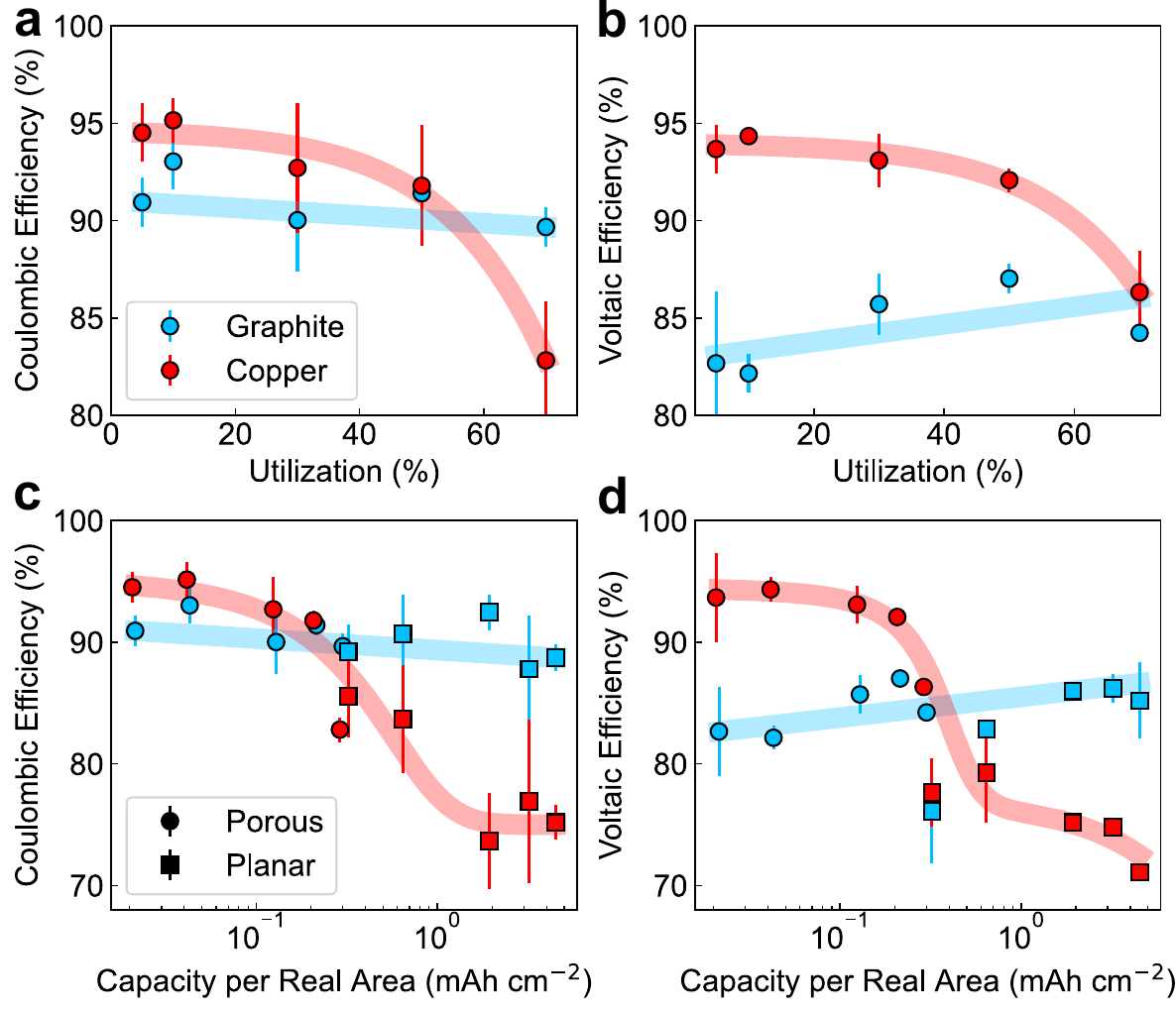}
  \caption{Coulombic efficiency \tb{a} and voltaic efficiency \tb{b} of pouch cell cycling over 500 hours as a function of utilization for both the porous graphite (blue) and porous Cu (red) substrates. Coulombic efficiency \tb{c} and voltaic efficiency \tb{d} of porous (circles) and planar (squares) substrate pouch cells over 500 hours of cycling as a function of plated Sn capacity per real surface area. Each condition is represented by triplicate cells, error bars correspond to the standard deviation between replicate cells, and shaded curves are guides to the eye.}
  \label{fig:fig5}
\end{figure}

\section{Discussion} \label{sec:conclusions}


Our \textit{operando} microscopy provides a detailed mechanistic understanding of Sn metal anode behavior as a function of both substrate chemistry and utilization. 
We demonstrated that Cu is a highly reactive substrate that alloys with Sn over the duration of battery operation, forming the semi-irreversible intermetallic \im{}. 
This strong interfacial interaction governs the morphological evolution of deposited Sn, modulates plating overpotential, and ultimately compromises reversibility at high plated capacity and cycle number. 
In contrast, graphite behaves as a comparatively inert substrate, enabling a single-stage deposition mechanism in which discrete Sn islands nucleate and grow independently. 
When coupled with the disproportionation reaction that elevates discharge voltage at higher plated capacities, this inert interfacial environment allows graphite anodes to sustain excellent performance at high areal plated capacities and high utilizations.

From these findings, we propose several broader characterization principles for metal deposition anodes in both aqueous and non-aqueous systems. 
First, \textit{operando} characterization over extended time periods is essential. Metal plating and stripping involves highly dynamic interfacial and morphological processes that cannot be reliably inferred from \textit{ex-situ} measurements and change over the course of battery state of charge and lifetime. 
Meaningful insight requires a robust \textit{operando} cell architecture that faithfully replicates the electrochemical behavior of control devices and a stable imaging platform to evaluate the system over many hours and cycles. 
In this work, the same pouch cell configuration was used for both electrochemical evaluation and \textit{operando} imaging, ensuring direct correspondence between mechanistic observations and practical performance.
This methodology allowed us to evaluate the evolution of the electrode-electrolyte interface over hundreds of hours and cycles, directly observing its degradation and impact on performance.

Second, evaluation at high utilization is critical, with both growth mechanisms and electrochemical stability evolving with increased plated capacity. 
Performance data collected only at low utilization risks obscuring degradation pathways that dominate under realistic operating regimes.
Here, we observe that under high areal plated capacities, Cu undergoes severe degradation, whereas graphite maintains stable performance, contrary to the conclusions of prior work that only evaluated Cu substrates under low utilizations and limited time-scales.

Collectively, our high-utilization, \textit{operando}-driven framework establishes a broadly applicable strategy for evaluating metal anode chemistries. 
We have shown that \textit{operando} microscopy reveals mechanistic underpinnings of device failure across their operational lifetime, and anticipate that this methodology will accelerate research and development of a wide range of electrochemical systems.

\section{Methods} \label{sec:methods}

\subsection{Materials}
Potassium stannate trihydrate (K$_2$SnO$_3$·3H$_2$O, 99.9\%) and potassium hydroxide (KOH, 99.99\%) were purchased from Sigma-Aldrich. MilliQ ultrapure water was produced from deionized (DI) water using a Simplicity water purification system (Millipore). Bilayer plastic pouches (SealPak 402-24, Kapak, 63.5 $\mu$m) were purchased from VWR. Graphite felt (G280A, AvCarb, 2.8 mm) was purchased from Fuel Cell Store. Planar graphite (Tgon 805 Electrically Conductive Graphite) was purchased from Digikey. Porous Cu foam (MF-Cu16Fom, 99.9\%) was purchased from MTI Corporation. Planar Cu (110, ASTM B152, 0.01" thick) and planar Nickel (200-201, ASTM B160, 9707K71, 25.4 $\mu$m) were purchased from McMaster-Carr. Polyolefin nonwoven separators (0.25 mm) were purchased from Freudenberg Performance Materials. Anion exchange membranes (Selemion ASVN) were purchased from Bellex International and stored in DI water prior to use.

\subsection{Device fabrication}

Plastic pouches were assembled by heat sealing two nickel tabs between two layers of the SealPak bilayer plastic material with a hot melt adhesive and an impulse heat sealer. Ni(OH)$_2$ cathodes were prepared by discharging Eneloop rechargeable nickel metal-hydride AA batteries (BK-3MCCA16FA, Panasonic) at 50 mA to 1.2 V before battery disassembly. The cathode rolls were harvested, flattened with a press and cut into  squares (1.5$\times$1.5 cm$^2$ for the porous graphite felt cells and 2$\times$2 cm$^2$ for all other cell formats). Selemion ASVN monovalent anion exchange membranes were soaked in MilliQ water and then cut into 3$\times$3 cm$^2$ squares. Graphite felt was cut into 1$\times$1 cm$^2$ squares while the porous Cu foam was first flattened with a press before being cut into 2$\times$2 cm$^2$. Planar graphite and planar Cu substrates, and nonwoven separators were also cut into 2$\times$2 cm$^2$ squares. 

Device stacks were assembled in the pouch as Ni(OH)$_2$/Selemion AEM/substrate for the porous cells and Ni(OH)$_2$/Selemion AEM/nonwoven separator/substrate for the planar substrate cells. Three sides of the pouches were heat sealed at 140 \degree C (MSK-115-III, MTI) including sealing the separator directly into the pouch to avoid leakage. The anolyte was prepared by dissolving 0.08 mol (4.49 g) KOH and 0.48 mol (143.5 g) \ksn{} in a 200 mL volumetric flask, adding DI water and mixing until the solution meniscus was at the calibration mark, resulting in an electrolyte with 0.4 M KOH and 2.4 M \ksn{}. The catholyte was created by dissolving 0.14 mol (7.86 g) of KOH in a 50 mL volumetric flask, adding deionized water and mixing until the solution meniscus was at the calibration mark to make a 2.8 M KOH solution. The catholyte and anolyte were micropipetted on each side of the AEM respectively. For all cells, 30 $\mu$l of catholyte was used. 200 $\mu$l and 45 $\mu$l of anolyte was used for the porous graphite and porous Cu cells respectively, ensuring that the amount of Sn per unit real area matched for both cell types. 100 $\mu$l of anolyte was used for both planar cell types. The fourth side of the pouch was vacuum sealed at 140 \degree C.

\subsection{Electrochemical Testing}

For pouch cell galvanostatic cycling, pouch cells were placed into a temperature chamber held at 30 \degree C (IC-150R, IncuMax) and connected to an external battery cycler (BCS-805, BioLogic). A constant current of 0.5 mA was applied to all cells during charge and discharge. For charge, capacity cutoffs were generally used that were based on percentage of utilization of Sn in the electrolyte, but an upper cutoff voltage of 2.1 V was applied for safety to protect for the edge case of excessive gas generation. Cells were then discharged to a voltage cutoff of 0.2 V for the graphite case and 0.8 V for the Cu case (to prevent surface oxidation). For over-discharge experiments, the Cu was also allowed to discharge to 0.2 V. 

\subsection{\textit{Operando} Optical Microscopy}

Optical microscopy experiments were performed on a home-built, wide-field inverted microscope loosely based on the design by Ortega Arroyo et al. \cite{ortegaarroyoInterferometricScatteringMicroscopy2016,merryweatherOperandoOpticalTracking2021}. Light from an LED source (white - MNWHLP1, red - M625L4, Thorlabs) was collimated and passed through a polarizing beam splitter such that the light was linearly polarized. The beam was then focused onto the back focal plane of the objective lens using an achromatic doublet lens with a 300 mm focal length and a visible anti-reflection coating (AC254-300-A, Thorlabs). Before the beam reached the objective, the light was passed through a quarter-wave plate to turn the linearly polarized light into circularly polarized light. The light was then collimated by the objective lens (CFI60 Plan Fluor 20x 0.75 NA Multi-Immersion Objective Lens, Nikon) before reaching the sample. Immersion oil was used to prevent reflections from the rough plastic pouch surface from reaching the camera. The light was reflected/scattered by the sample before being recollected by the objective lens and passed once again through the quarter wave plate to turn the circular polarization back into linear polarization but with the polarization rotated by 90\degree\ from its original orientation. As such, when the beam was incident on the polarizing beam splitter a second time, it was reflected towards the camera instead of being transmitted. The same achromatic doublet lens was also used as a tube lens to focus the light onto the camera (Grasshopper GS3-U3-23S6C-C, FLIR).

The objective lens was mounted directly onto a rigid breadboard. The sample was mounted with a custom 3D printed bracket onto a piezo-driven xyz stage consisting of three monodirectional stages mounted orthogonally on top of one another (ECSx5050/Al/NUM/RT, Attocube). The sample was moved into the focal plane using the piezoelectric stage and was kept in the focal plane using an image contrast-based feedback loop similar to autofocusing in some phone cameras. For RGB images, the image was converted to grayscale for subsequent computation. Briefly, a region of the image was chosen to keep in focus. This portion of the image ($I$) was median filtered with a 3$\times$3 pixel kernel to ensure there were no outlier hot/cold pixels. The Laplacian ($\nabla^2I$ or $\frac{\partial^2 I}{\partial x^2} + \frac{\partial^2 I}{\partial y^2}$) of the region was then calculated, providing a measure of the two dimensional curvature at each point of the region. This function works effectively as an edge detection algorithm - it will be close to zero in regions of the image where the intensity is either constant or increases very smoothly/linearly, whereas it will be large in absolute value where there are sharp changes. Images in focus are defined by sharply defined edges of objects, while images out of focus are smeared out and lack well defined edges. Focused images will therefore have the largest variance in the Laplacian. We continuously monitor the Laplacian, and if it drops below a threshold value relative to its maximum, the stage takes incremental steps up the gradient of increasing contrast until it finds the maximum value. This enables the sample to remain in focus over the extended duration of these experiments (hundreds of hours) with minimal impact from pouch cell factors such as Joule heating, swelling from gas generation, or external factors such as stage/table vibration or temperature changes in the room. 

Identical pouch cells to those used in the main electrochemistry experiments were used for the \textit{operando} optical measurements. Cells were cycled galvanostatically using a portable potentiostat (SP-300, BioLogic). The cells were also cycled to the relevant capacity cutoff, with the only correction being a reduced upper cutoff voltage of 1.82 V. When cells approached this voltage, HER could be directly observed, as a gas bubble would form between the pouch cell material and the electrolyte, obscuring the electrode surface. The lower voltage was set to avoid this edge case. 

\subsection{X-ray Diffraction}
\textit{Ex-situ} grazing incidence X-ray diffraction measurements (GIXRD) of the planar electrodes were carried out using a PANalytical Empyrean X-ray diffractometer equipped with a Cu source ($\lambda$ = 1.54 \AA{}). The incident angle was fixed at 0.5\degree{} to enhance the signal from the surface layers. The detector was scanned over the two theta range from 25\degree{} to 80\degree{} with a 0.01\degree{} step size.

\subsection{Electron Microscopy}
TEM lamella were prepared using a Thermo Fisher Scientific Helios Hydra DualBeam plasma focused ion beam-scanning electron microscope (FIB-SEM) operated at 30 kV with an Ar$^+$ plasma source.
Scanning transmission electron microscopy (STEM) was performed on a Thermo Fisher Scientific Spectra 300 monochromated, double-corrected scanning transmission electron microscope operated at 300 keV. Elemental analysis was performed using a SuperX energy dispersive spectroscopy (EDS) detector.
Scanning electron microscopy (SEM) was performed on a FEI Magellan 400 XHR Scanning Electron Microscope.

\section{Data Availability}
The data supporting the findings of this work will be made available before publication at the repository: DOI:XXXXXXX.

\section{Acknowledgements}
K.F. acknowledges funding and support from the Stanford Energy Postdoctoral Fellowship and the Precourt Institute for Energy.
Part of this work was performed at nano@stanford RRID:SCR026695. K.F. acknowledges the Lab64 maker space for access to fabrication tools. The authors acknowledge J. Tyler Mefford for his foundational ideas related to this study and Yan-Kai Tzeng for helpful discussion. The authors thank Adam Marks, Edward Mu, and Molly Corr for helpful comments on the manuscript.

\section{Author Contributions}
S.K.C. and K.F. conceived the ideas in the work and designed the research with the supervision of D.N.C. and W.C.C. S.K.C., K.F. and W.T. built the electrochemical cells and performed the electrochemical characterization. K.F. designed and built the operando optical microscopy setup and performed the measurements with assistance from S.K.C. and W.T. S.K.C. and D.Y. performed the SEM experiments. D.Y. performed the TEM measurements under the supervision of C.O. K.H. performed the GIXRD measurements with assistance from S.K.C. and K.F. J.W. contributed to the conception and mechanistic interpretation of the work. S.K.C. and K.F. wrote the manuscript with input from D.N.C. and W.C.C. All authors provided comments and feedback on the manuscript.

\section{Competing Interests}
The authors declare no competing interests.



\bibliography{ref_kyle}

\end{document}